\documentclass[12pt,a4paper]{iopart}
\usepackage{iopams}
\usepackage{graphicx}
\usepackage{epstopdf}
\usepackage{xspace}
\usepackage{color}
\usepackage{hyperref}

\begin{document}

\title[The lifetime problem of evaporating black holes: mutiny or 
resignation]{The lifetime problem of evaporating black holes: mutiny or 
resignation}

\author{Carlos~Barcel\'o$^1$, Ra\'ul~Carballo-Rubio$^1$, Luis~J.~Garay$^{2,3}$, 
and Gil Jannes$^{4}$}

\address{$^1$ Instituto de Astrof\'{\i}sica de Andaluc\'{\i}a (IAA-CSIC), 
Glorieta de la Astronom\'{\i}a, 18008 Granada, Spain}
\address{$^2$ Departamento de F\'{\i}sica Te\'orica II, Universidad Complutense 
de Madrid, 28040 Madrid, Spain}
\address{$^3$ Instituto de Estructura de la Materia (IEM-CSIC), Serrano 121, 
28006 Madrid, Spain}
\address{$^4$ Modelling \& Numerical Simulation Group, Universidad Carlos III 
de Madrid, Avda. de la Universidad 30, 28911 Legan\'es, Spain}

\eads{\mailto{carlos@iaa.es}, \mailto{raulc@iaa.es}, 
\mailto{luisj.garay@ucm.es}, \mailto{gil.jannes@uc3m.es}}

\begin{abstract}

It is logically possible that regularly evaporating black holes exist in 
nature. In fact, the prevalent theoretical view is that these are indeed the 
real objects 
behind the curtain in astrophysical scenarios. There are several proposals for 
regularizing the classical singularity of black holes so that their 
formation and evaporation do not lead to information-loss problems. One 
characteristic is shared by most of these proposals: these regularly 
evaporating black holes present long-lived trapping horizons, with absolutely 
enormous evaporation lifetimes in whatever measure. Guided by the discomfort 
with these enormous and thus inaccessible lifetimes, we elaborate here on an 
alternative regularization of the classical singularity, previously proposed by the authors in an emergent gravity framework, which leads to a completely 
different scenario. In our scheme the collapse of a stellar object would result in a genuine time-symmetric bounce, which in geometrical terms amounts to the 
connection of a black-hole geometry with a white-hole geometry in a regular 
manner. The two most differential characteristics of this proposal are: i) the 
complete bouncing geometry is a solution of standard classical general relativity 
everywhere except in a transient region that necessarily extends beyond the 
gravitational radius associated with the total mass of the collapsing object; and ii) the duration of the bounce as seen by external observers is very brief 
(fractions of milliseconds for neutron-star-like collapses). This scenario 
motivates the search for new forms of stellar equilibrium different from black 
holes. In a brief epilogue we compare our proposal with a similar geometrical 
setting recently proposed by Haggard and Rovelli.       

\end{abstract}


\tableofcontents

\section{Introduction}
\label{Sec:introduction}

Standard general relativity holds the possibility of forming black holes under 
mild hypotheses requiring positivity of the energy. These classical black 
holes would possess as defining characteristics an event horizon and some 
internal singularity. They would be absolutely inert objects, the ultimate end 
point of collapse, the dead state of stellar physics. However, if only by the 
presence of singularities, researchers have been strongly inclined to believe 
that there should exist a deeper-layer theory free of these singularities. The 
first suspect missing ingredient is of course quantum mechanics.

When trying to build, at least partially, coherent scenarios in which the 
collapse of matter is affected by the very quantum nature of matter, the 
previous situation significantly changes. Robust semiclassical calculations 
tell us that black holes could not be absolutely stationary, they would have to 
evaporate by emitting Hawking quanta~\cite{Hawking1975}. The question then is: 
Are these evaporating black holes really black holes in the sense of having an 
absolute event horizon? Or a directly related question: Is there some 
information lost in a complete evaporation process? These questions have been a 
matter of controversy, and a strong driving force for theoretical development, 
for about 40 years (the controversy started with~\cite{Hawking1976}; a personal 
recount of its history can be found in~\cite{Susskind2008}). However, nowadays 
even S. Hawking concedes that the most reasonable solution is that the 
classical internal would-be singularity is quantum-mechanically regularized in 
such a way that no strict event horizon would ever form, only long-lived 
trapping horizons~\cite{Hawking2014} (see \cite{Visser2014} for a recent discussion on the observational differences between event and trapping horizons). Therefore, under this view, strictly 
speaking black holes ({\it i.e.} causally disconnected regions) would not exist 
in nature. However, owing to the similarity of these quantum-corrected objects 
with classical black holes, they usually keep the  name ``black hole''. Here we 
will always call them generically regularly evaporating black holes (REBHs) to distinguish 
them from their classical cousins.  

Thus, the prevalent view is that REBHs indeed form in astrophysical scenarios 
and that, by themselves ({\it i.e.} forgetting their gravitational interaction 
with the surrounding matter), they would remain almost inert for very many 
Hubble times except for a tiny evaporative effect that eventually would make 
them disappear. The evaporation rate of stellar-mass objects would be so slow, 
$10^{67}$ years to halve the size of a Solar mass object, that for all 
practical purposes they could be considered stationary. Whereas the distinction 
between a classical BH and a REBH is of fundamental interest on purely 
theoretical grounds, it is almost certainly irrelevant for all astrophysical 
purposes (disclaimer: nature might still surprise us with the existence of 
primordial black holes~\cite{Hawking1974}; see e.g.~\cite{Carr2010} for some 
recent constraints on their existence).

Different evaporation scenarios have been put forward to accommodate solutions 
to the information problem. As a sampling of the vast literature on the subject, let us mention 
the complementarity approach of Susskind~\cite{Susskind1993}, the 
fuzzball proposal of Mathur~\cite{Mathur2005}, the condensed state of 
Dvali~\cite{Dvali2014}, the Giddings remnant scenario~\cite{Giddings1992} and 
the loop-quantum-cosmology-inspired scenario of Ashtekar and 
Bojowald~\cite{Ashtekar2005}. More recently, also inspired by ideas from loop 
quantum cosmology, Rovelli and Vidotto~\cite{Rovelli2014} have put forward an 
evaporation model such that when the collapsing matter approaches the classical 
would-be singularity, it undergoes a kind of quantum bounce (there is a similar 
proposal by Bambi {\it et al.}~\cite{Bambi2014} based on asymptotic freedom). However, 
what we want to emphasize here is that all these models share a common feature: 
All of them contemplate REBHs as essentially hollow and (almost) stationary 
regions of spacetime; {\it i.e.} all of them share the very same 
quasistationary view of extremely slowly evaporating trapping horizons.

\subsection{The lifetime problem}
\label{Subsec:scenario}

REBHs have an enormous lifetime, in whatever measure. And this becomes even 
worse when considering that astrophysical black holes are actually on average 
growing and not yet evaporating because their Hawking temperature is smaller 
than the $\sim$ 3K of the Cosmic Microwave Background. This implies that any external experimental test ({\it i.e.} whose 
result is available to observers outside black holes) of the precise way in 
which the evaporation proceeds and, in turn, of the very nature of the internal 
regularization actually operating, would be almost certainly beyond the reach of human kind. 
Before entering into a resignation mood and accepting this as one of those 
``such is life'' cases, we think it worth to analyze whether there is still an 
escape route. 

The lifetime problem is rarely highlighted in the literature. As a notable 
exception, Penrose has shown explicit signs of discomfort with this issue, 
suggesting an interesting solution~\cite{Penrose2011}. He argues that in the 
long run, all massive particles might disappear, leaving a conformally 
invariant world. In a conformally invariant world duration has no meaning, so 
one cannot worry about duration of physical processes and more specifically of 
the evaporation process.

Here, we want to draw attention to a completely different take on this problem 
and in turn on the internal regularization issue. In simple geometrical terms 
it amounts to regularly connecting a black hole geometry with a white-hole 
geometry and hence producing a bouncing geometry with a {\em very short 
bouncing time as seen from external observers} (of the order of fractions of a millisecond 
for neutron-star-like initial configurations). In terms of producing a 
bounce, our proposal goes further than that of Rovelli-Vidotto by regularizing 
the classical would-be singularity by a genuine time-symmetric bounce. 

Three of the authors of this paper entertained this idea years ago while 
discussing the possibility of constructing a coherent theory of emergent 
gravity from condensed-matter-like systems~\cite{Barcelo2010a}. The main 
idea of that paper was that the first ``quantum gravity'' effect could be the 
switching-off of gravity, a situation that would be prompted when Planck 
densities are at stake. Here we will show in general terms that the 
regularization we propose only requires that the causality of the next layer 
beyond general relativity, prompted at high energies, is different from the 
effective causality described by classical general relativity, and all this 
independently of the specific nature of this new layer.

Let us mention here that the idea of connecting a black hole with a white hole 
was also argued for in an earlier paper by H\'aj\'i\v{c}ek and 
Kiefer~\cite{Hajicek2001}. They obtained such a configuration by quantizing a 
system consisting of a spherical null shell of matter. They did not mention the 
lifetime problem though, but argued that the resulting object could still be 
locally similar to a black hole. However, as far as we can see, their result 
might suggest otherwise and could be taken as further support, from a 
different angle, for our proposal. During the writing of this paper, which is a 
more detailed version of the essay~\cite{Barcelo2014e}, a new paper by Haggard 
and Rovelli~\cite{Haggard2014} has appeared in the 
arXiv, which shares several 
ingredients with our proposal. In an epilogue we shall briefly discuss the 
similarities and dissimilarities between our approaches.

Before describing in detail our actual proposal let us recall the current 
situation in the white-hole district.       

\subsection{The white-hole district}
\label{Subsec:white-hole}

The simplest models of collapse are spherically symmetric and we will 
concentrate on them. Birkhoff's theorem is often thought of as a uniqueness 
result. This is certainly the case for the vacuum geometry outside a spherical 
distribution of matter exterior to the Schwarzschild radius. However, for 
vacuum geometries inside the Schwarzschild radius this is not true. Birkhoff's 
theorem asserts that any vacuum patch must be locally equivalent to a patch of 
the maximally extended Kruskal solution~\cite{Hawking1973}. The vacuum geometry 
outside a spherical distribution of matter but still inside the Schwarzschild 
radius can thus be of black-hole or of white-hole type. The Schwarzschild 
interior regions are dynamical and can be either contracting towards a 
singularity (black-hole district) or expanding from a singularity (white-hole 
district). Birkhoff's theorem does not tell us which internal geometry is the 
appropriate one.
 
Within classical general relativity stellar objects eventually collapse to form 
black holes and future singularities. This is arguably the main reason why the 
white-hole region is considered as non-physical and is almost forgotten in 
what, unsurprisingly, happens to be known as ``black hole physics''. The 
strange oblivion of white holes was highlighted many years ago in a sharp 
article by Narlikar, Appa Rao and Dadhich~\cite{Narlikar1974}.

In Newtonian gravity one can perfectly imagine arbitrarily small balls with 
arbitrarily large masses. One can design non-collision scattering processes 
with arbitrarily small impact parameter. These scatterings are characterized by 
an acceleration phase (falling into the gravitational potential) followed by a 
deceleration phase (climbing the gravitational potential). The entire 
scattering is perfectly time symmetric. By considering the balls as hard 
spheres subject to elastic collisions or non-interacting with each other when coming into superposition ({\it i.e.} transparent to each other), one could even make a precise head-on collision. This process will also consist of 
the previous two phases. 

Now, general-relativistic situations in which a lump of matter is climbing a 
gravitational potential can easily be found: one can just think of throwing a 
stone upwards over our heads. One can also think of the following idealized 
thought experiment. Two equal balls with sufficiently small mass-to-radius 
ratio are thrown directly towards each other. Imagine that they are transparent 
to each other (for simplicity one could imagine them to be internally rigid to a first 
approximation and forget about changes in their form). In the case in 
which no horizon forms ({\it i.e.} sufficiently low kinetic energy in the 
collision), if one neglects the gravitational wave emission due to the 
encounter, the resulting trajectories will be perfectly time symmetric, again 
with acceleration and deceleration phases. Within standard classical general 
relativity the situation radically changes if a horizon forms. In a collision 
with sufficiently high kinetic energy the two balls in the previous example will 
form a horizon. The time-symmetric deceleration phase will disappear and be
substituted by a black hole remnant. Somewhat surprisingly, although white-hole 
solutions exist mathematically, in supposedly realistic situations one never 
encounters potential-climbing cases beyond the Schwarzschild radius. 

The singular Coulomb attraction of Newtonian gravity needs a regularization when dealing with finite size particles (for point-like particles the probability of collisions is a set of null measure). As mentioned, this regularization could be for example to assume elastic hard-core collisions or transparency. In reality this pathological behaviour at short distance of Newtonian gravity is maintained by general relativity but in a subtle manner: hidden by horizons. The problem in general relativity is that simple regularizations as the previous ones appear difficult to implement straightforwardly. This paper is about a short-distance regularization proposal in the simplest situation one can think about. It is because of this short-distance nature that we can think of the regularization as being quantum in nature 
in generic terms.

The fading into oblivion of the white-hole district is arguably the strongest 
departure of general relativity from Newtonian gravity. There seems to be a 
prejudice against exploring the possibility and implications of really having 
genuine bouncing solutions. As we will see, there are indeed good reasons for 
this prejudice to exist, but also for taking the bold leap of exploring beyond 
it.
 
What would happen if the collapsing matter underwent a perfectly genuine 
time-symmetric bounce when reaching Planck densities? As described above, in Newtonian gravity hard-core or transparent regularizations will lead to genuine bouncing solutions. However, within the evaporation model of 
Rovelli-Vidotto the bouncing is asymmetric in time. Let us explain what we 
mean by that. The collapse of matter beyond its Schwarzschild radius produces a 
trapped region. When the collapsing matter reaches Planck densities it 
encounters a quantum repulsion that makes the collapse halt and bounce. Then, 
the way out of the bouncing encounters the opposition of the previously bended 
lightcones in the trapped region. As the bouncing progresses outwards, 
accompanied by the evaporation of the structure \`{a} la Hawking, these 
lightcones slowly recover their unbent orientations before the collapse. 
Finally the outer and inner horizons coalesce and the trapped region 
disappears. The expanding phase is intrinsically distinct from the collapsing 
phase: for instance there is no inner trapped surface whatsoever in the 
geometry, 
nothing related with a while-hole geometry. In the next section we will see 
more clearly the difference between an asymmetric bouncing and a symmetric one 
and its connection with the lifetime problem.

\section{A genuine bouncing scenario}
\label{Sec:scenario}

Basically we want to analyze in generic qualitative terms what would happen if the collapsing matter, upon 
reaching Planck density, slowed down and bounced back, connecting with the 
time-reversed geometry associated with a white-hole spacetime. Our metric contains a number of unknown parameters that would depend on the details of the underlying theory. However, it also possesses some marked characteristics that clearly distinguishes it from other proposals.
After describing these metric characteristics, the rest of the paper will be devoted to a discussion of the paradigm shift this regularization entails. First we will describe the bouncing geometry both from an explicitly time-symmetric point of 
view and from an external point of view. These two views amount to two different 
choices of coordinates. At the end of this section, in 
Subsec.~\ref{Subsec:hypothesis}, we will discuss one crucial characteristic 
that any deeper layer underneath general relativity should have for this 
scenario to be plausible. 

\subsection{Explicitly time-symmetric view and external view}
\label{Subsec:explicitly-time-symmetric}

Let us build a simple specific geometry representative of a time-symmetric 
bouncing regularization of a collapsing star. We are going to concentrate on 
spherically symmetric collapses and avoid additional complications at this 
stage. Classically, a spherically-symmetric collapse does not produce 
dissipation in the form of gravitational radiation. If we took into account 
quantum corrections, there would indeed exist some dissipation. However, in the 
limit of very large masses, this quantum radiation could in principle be made  
as small as desired. The time-symmetric geometry we are going to present is a 
reasonable dissipationless approximation to more realistic, dissipative 
situations.

\begin{figure}[h]%
\vbox{ \hfil  \includegraphics[width=0.67\textwidth]{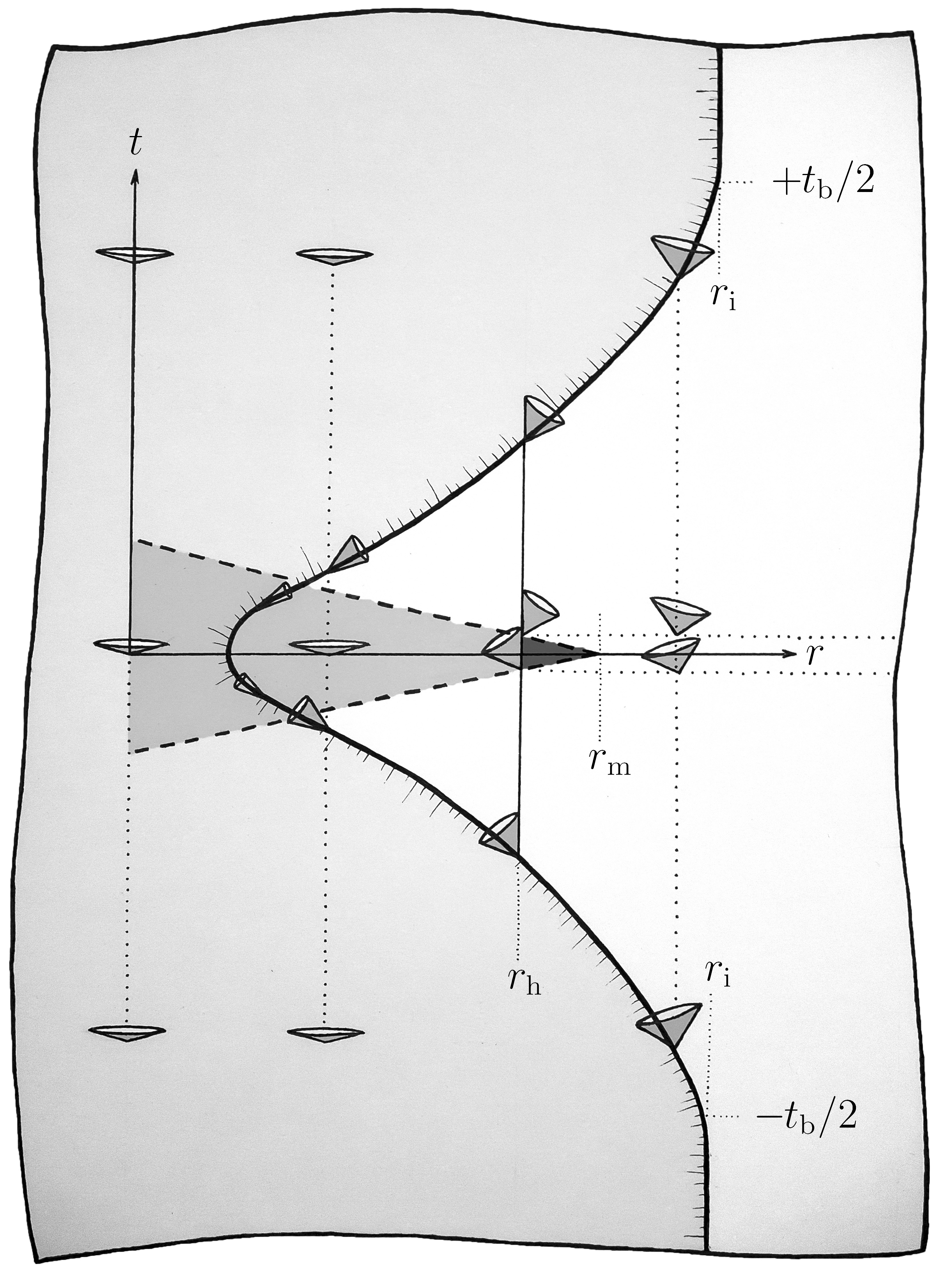}\hfil}
\bigskip%
\caption{The figure represents the collapse and time-symmetric bounce of a 
stellar object in our proposal (the thick line). The past thick dashed line from $r=0$ to $r_{\rm m}$ marks the boundary where the non-standard gravitational effects start to occur. In all the external white region the metric is Schwarzschild. In the region 
between the two thick dashed lines (which extends outside the stellar matter itself) the metric is not 
Schwarzschild, including the small dark grey triangle outside the Schwarzschild 
radius $r_{\rm h}$. The drawing tries to capture the general features of any interpolating geometry. The slope of the almost Minkowskian cones close to the origin has been taken larger than the usual 45 degrees to cope with a convenient and explicit time-symmetric drawing.}
\label{Fig:bh-wh-diagram}%
\end{figure}%

Figure~\ref{Fig:bh-wh-diagram} represents a bouncing geometry associated with a 
spherically-symmetric homogeneous dust star (Oppenheimer-Snyder 
model~\cite{Oppenheimer1939}) collapsing from an initial radius $r_{\rm i}$ far 
above its Schwarzschild radius $r_{\rm i} > 2M$. To depict this 
geometry we assume that before and after the bounce the geometry is static, {\it 
i.e.}: we concentrate on the geometry associated with a single bounce. Let us 
describe this implosion-explosion metric in generalized Painlev\'e-Gullstrand 
coordinates~\cite{Kanai2011}. 
These coordinates are adapted to observers attached to the stellar 
body. In~\cite{Barcelo2014e} we considered the slightly simpler case of a 
collapse starting from rest at infinity, so there we used the standard 
Painlev\'e-Gullstrand coordinates. Here we consider that the collapse starts 
from an initial radius $r_{\rm i}$. Therefore, we shall write the metric as~\cite{Kanai2011} 
\begin{eqnarray}
ds^2=-c^2 dt^2 + {1 \over \epsilon^2}(dr - v dt)^2+r^2 d\Omega_2^2.
\label{generalized-acoustic-metric}
\end{eqnarray}
Here, the three functions $c(t,r),\epsilon(t,r),v(t,r)$ will now be given patch 
by patch. The collapse of the star begins from rest at the initial radius 
$r_{\rm i}$ and is represented by the trajectory of the star's surface 
$r_s(t)$. The arbitrary zero of time is chosen such that the collapse 
starts at $t=-t_{\rm b}/2$, where
\begin{equation}
t_{\rm b}:=2\pi\left({r^3_{\rm i} \over 
8M}\right)^{1/2}\label{eq:tbounce}
\end{equation}
is twice the classical collapsing time (see below). For $t<-t_{\rm b}/2$ the metric is exactly 
Schwarzschild outside $r_{\rm i}$. Inside $r_{\rm i}$, $c=1$, $v=0$ and 
$\epsilon=1-2Mr^2/r_{\rm i}^3$. This corresponds to a star of homogeneous 
density, maintained static by an appropriate internal pressure.
Once the collapse has started, the three coordinate patches we will use are 
$0<r<r_s(t)$, $r_s(t)<r<r_{\rm i}$ and $r_{\rm i}<r<+\infty$. 
Let us first consider the region between the hypersurface $t=-t_{\rm 
b}/2$ and the one defined by the thick dashed line from $r=0$ 
up to $r=r_{\rm m}$ continued by the $t=0$ line from $r=r_{\rm m}$ up to 
infinity (see Fig.~\ref{Fig:bh-wh-diagram}; note that $r_{\rm m}$ is by definition the farthest radius reached by the non-standard transient region, see below). In that region the metric functions are:
\begin{eqnarray}
c^2:= \left\{
\begin{array}{l}
\left(1-{2M \over r}\right)/\left(1-{2M \over r_{\rm i}}\right),~~~ r>r_{\rm 
i}, \\
1,~~~ r_s(t)<r<r_{\rm i}, \\
1,~~~ 0<r<r_s(t), 
\end{array}
\right.
\label{c-profile}
\end{eqnarray}
\begin{eqnarray}
\epsilon^2:= \left\{
\begin{array}{l}
1-{2M \over r},~~~ r>r_{\rm i}, \\
1-{2M \over r_{\rm i}},~~~r_s(t)<r<r_{\rm i}, \\
1-{2M \over r_{\rm i}}\left({r \over r_s(t)}\right)^2,~~~0<r<r_s(t),
\end{array}
\right.
\label{epsilon-profile}
\end{eqnarray}
\begin{eqnarray}
v =v_{\rm I}:= \left\{
\begin{array}{l}
0,~~~r>r_{\rm i}, \\
-\sqrt{{2M \over r}-{2M \over r_{\rm i}}},~~~r_s(t)<r<r_{\rm i}, \\
-\sqrt{{2M \over r_s(t)}-{2M \over r_{\rm i}}}{r \over r_s(t)},~~~ 0<r<r_s(t). 
\end{array}
\right.
\label{velocity-profile}
\end{eqnarray}
In this region the function $r_s(t)$ is the one calculated in standard 
general relativity for a collapsing homogeneous dust ball (the 
Oppenheimer-Snyder collapse):
\begin{eqnarray}
r_s(t)= {r_{\rm i} \over 2}(1 +\cos \eta),~~~~t= \frac{t_{\rm b}}{2\pi}(\eta +\sin \eta-\pi),
\end{eqnarray}
with $\eta \in [0,\pi-\varepsilon]$ and $\varepsilon$ a small positive 
constant. In the classical case in which the collapse forms a singularity, this 
same metric will continue until the singularity at $\eta=\pi$ is reached. 
However, in our proposal the singularity $r_s=0$ is not reached. Instead our 
regular trajectory $r_s(t)$ is constructed in the following way: i) take the 
previous singular collapsing function, {\it i.e.} the $r_s$ with $\eta \in 
[0,\pi]$, ii) continue it with its time reversal, also a classical solution 
associated now with matter appearing from the singularity, and iii) modify 
the singular $r_s=0$ part of the trajectory in a small interval (of order $\varepsilon$) around $t=0$ 
(in between the two thick dashed lines in Fig.~\ref{Fig:bh-wh-diagram}) so 
that the total trajectory represents a time-symmetric bounce. The departure 
from 
the general relativity behaviour will start at $\eta=\pi-\varepsilon$. 
In the region above the hypersurface delimited by the top thick dashed line from $r=0$ to $r=r_{\rm m}$ and continued by the $t=0$ line from $r=r_{\rm m}$ to infinity, the metric has the same form as before but only with $v=v_{\rm 
E}:=-v_{\rm I}$. It is the time-symmetric version (expanding 
phase) of the metric below the bottom thick dashed line, and so also a 
solution of the standard Einstein equations. Above $t_{\rm b}/2$ the 
metric is again static and thus equal to the one below $-t_{\rm b}/2$. The 
metric in the near triangular region between the two thick dashed lines 
extending up to $r_{\rm m}$ is non-standard. It is smooth and time-symmetric, 
interpolating the imploding and exploding patches. The $t=0$ slice represents 
the hyperplane of time-reversal symmetry. This region has two crucial 
characteristics: i) it extends beyond the classical gravitational radius, and ii) 
its temporal extension is very short.

Let us explain the reasons for these two characteristics. In the following discussion we 
will forget for a moment about the metric beyond $r_{\rm i}$, since the 
specific metric we are describing is always static and Schwarzschild 
in that region. Consider the region enclosed by the $r=r_{\rm h}$ vertical line to the 
left (the gravitational radius associated to the total mass of the object), by 
the  $r=r_{\rm i}$ vertical line to the right and by the two horizontal dashed 
lines above and below. In this region the light-cone structure smoothly turns 
towards its time-reversed version. This can be encoded in the previous metric 
Ansatz with an appropriately symmetrized function $v(t,r)=-v(-t,r)$ and smooth 
matchings at the boundary lines. One can always prescribe functions $v(t,r)$ 
and $c(t,r)$ for this region such that only in an open region containing the 
Schwarzschild radius the geometry is actually different from Schwarzschild's. 
In our metric representative we have taken this region to be the small grey 
triangle between the two thick dashed lines. Without more information 
about the bouncing processes we cannot determine the actual shape and size of this 
region, only that it should be a symmetric and continuous deformation of our 
representative such that it will always extend beyond the Schwarzschild 
radius.
Indeed, the change of coordinates    
\begin{eqnarray}
t'=t + \int_{r_{\rm i}}^r {2v(t,r) \over 1-v^2(t,r)/\epsilon^2(t,r)} dr,\qquad 
r'=r,
\end{eqnarray}
with $v(t,r)$ a function interpolating in time between $v_{\rm I}$ and 
$v_{\rm E}$, changes the metric (\ref{generalized-acoustic-metric}) from 
$v=v_{\rm I}$ to $v=v_{\rm E}$ without changing the geometry. 
These changes of coordinates would be ill-defined at the horizon itself, 
$|v|=\epsilon_{\rm i}$, with $\epsilon_{\rm i}=\left(1-2M / r_{\rm i} 
\right)^{1/2}$. Therefore, there must always be a region enclosing a portion of 
the Schwarzschild radius where the geometry must be different from 
Schwarzschild. Again, in the metric representative we are describing, we have 
taken this region to be the dark grey triangle. The necessary existence of a region 
with these characteristics is easy to understand. When connecting a 
black-hole with a white-hole geometry, the reversal of the light-cone 
tilting at the Schwarzschild radius is an essential physical process that mutates an 
initially marginally trapped surface to a non-trapping surface. On the 
other hand, outside the Schwarzschild radius there exist reversals of the 
light-cone bending that are just coordinate artefacts. We are using this 
property to describe the complete lightcone structure in explicitly 
time-symmetric terms.
 
\begin{figure}%
\vbox{ \hfil  \includegraphics[width=0.4\textwidth]{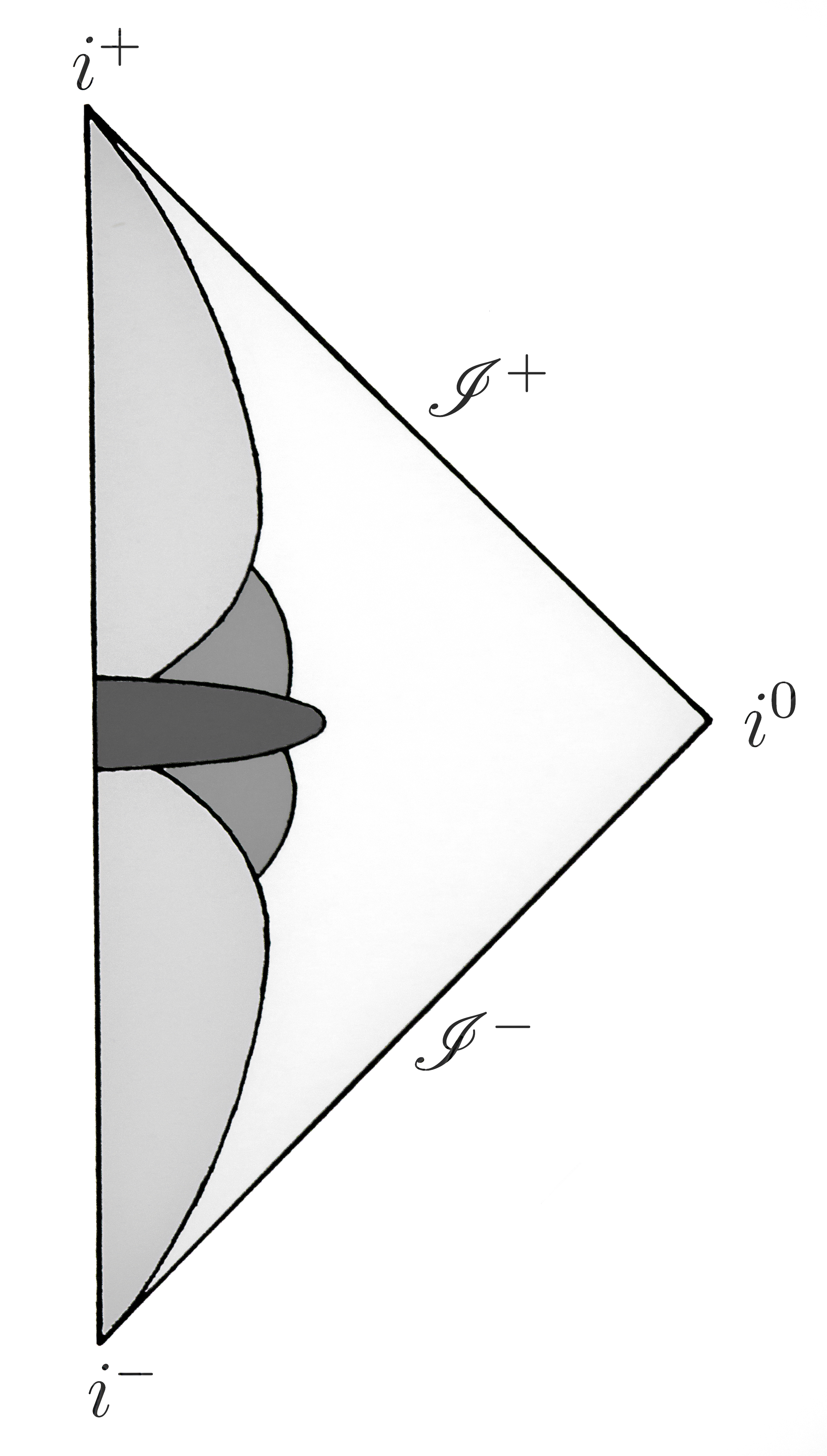}\hfil}
\bigskip%
\caption{The figure represents the Penrose diagram of the proposed geometry. Globally it has the same causality than Minkowski spacetime. Locally it has some peculiarities. The dark grey region represents a non-standard gravitational field, while the down and up grey regions are respectively regions with outer and inner trapped surfaces. The light grey regions on the left-hand side are those filled by matter.}
\label{Fig:penrose-diagram}%
\end{figure}

The Penrose diagram of the geometry just described was presented 
in~\cite{Barcelo2010a} and can be seen also here in 
Fig.~\ref{Fig:penrose-diagram}.
Causally this geometry is equivalent to Minkowski spacetime. Metrically 
speaking there are some differences: it contains a region of outer trapped 
surfaces and another region of inner trapped surfaces (we follow the convention of Hawking-Ellis \cite{Hawking1973}).  

From the perspective of an observer collapsing with the star, the entire process 
of collapse and bounce with respect to some reference initial position takes twice the 
free-fall collapsing time. By construction of the coordinate system, the 
proper time of the observer attached to the stellar structure is simply
the Painlev\'e-Gullstrand time, $d\tau=dt$. That is true because for that 
observer $dr-v\,dt=0$. For an initial, say, neutron star with 1.5 Solar masses and twice the Schwarzschild radius this time can be easily calculated from Eq. (\ref{eq:tbounce}) to be of the order of tenths of a millisecond. This same process seen by an observer 
always at rest at the initial position $r_{\rm i}$ takes {\em this very same 
time}. Indeed, for this observer $v=0$ so that $d\tau= dt$ is also its proper time 
parameter. The proper time separation between the two events, the start of the 
collapse and its return to the initial position, is the same for the surface-attached
observer and the one standing still at the initial radial position.
Seen by observers far outside the collapsing star it would take this time 
multiplied by the standard general-relativistic redshift factor, $(1-2M/ 
r_{\rm i})^{-1/2}$, associated with the reference position. This number will 
be of order unity for stars initially larger than a few times their Schwarzschild 
radius. Therefore, the lifespan of the complete bouncing process will be of the order of tenths 
of a millisecond also as seen from outside. Thus, these geometries do not exhibit 
long-lived trapped regions of any sort, only transient trapped regions.

As stated above, the expanding regime can be described by means of a patch of a white-hole geometry (excluding a region surrounding the past singularity). It is well known that there exist analyses concluding that white holes are unstable, so one might wonder whether this affects our proposal. To begin with, white holes (exact time reversals of black holes formed by collapse) exhibit a classical instability under the accretion of additional surrounding matter~\cite{Eardley1974,FrolovNovikov}. It has been shown that accretion could inhibit their explosion in the long run while for short time intervals the explosion can proceed unaffected. More detailed analyses of this instability~\cite{Barrabes1993,Ori1994} point out that the final release of energy depends essentially on the ratio between the mass-energy associated with the white hole and that of the accreting matter. These analyses are perfectly consistent with our model of ultrarapid expansion. Our simplest initial model does not contain additional accreting matter. If one had some matter climbing its gravitational potential (our white-hole phase) and it encountered some additional matter going inwards, the rate of expansion would decrease after crossing. The details of this process will depend on the place in which these two matter components meet each other and on their mass-energy ratio (in an idealized case in which they encounter each other without initiating additional non-gravitational effects). A detailed case-by-case analysis of this type of processes is beyond the scope of the present paper. Beyond these accreting processes, the formation of a white hole horizon in itself entails the appearance of a dynamical instability due to their blue-shifting behaviour~\cite{Szekeres1973,Lake1976}. This implies that white holes cannot last long. Again, this instability does not interfere with our model. Quite the opposite: it is in agreement with and can be taken as an additional physical motivation for our fast elimination of the outer trapped region. For completeness, let us mention that white holes would also suffer from a quantum-mechanical instability~\cite{Zeldovich1974}; the same previous comment applies to this spontaneous instability.

Let us go back to the model just presented but described from a different point of view. Imagine that one were to monitor with high time 
resolution this time-symmetric collapse from the reference initial position, 
which is assumed to be sufficiently far outside the Schwarzschild radius. One 
sets up two synchronized clocks at this initial position before the collapse. 
Then one clock is left to follow the collapsing structure and the other is kept 
at rest in the reference position. By looking with a telescope to the ticks of 
the clock falling with the star, one would see that in the collapsing phase the 
ticks slow down progressively. However, at some point they start to speed up in 
such a way that when the two clocks are finally back together they show 
precisely the same time.

For an external observer the collapsing phase will last longer than the 
expanding phase (this asymmetric view was the one presented 
in~\cite{Barcelo2010a}). Thus, he might not realize the time-symmetric 
character of the process. However, as said before, the total duration 
experienced by both observers will be equal and of the order of tenths of a 
millisecond (for neutron-star-like collapses).

In this paper we are interested in the general characteristics of the 
bouncing metric we are proposing. Its precise shape will depend on the details 
of the specific regularization process at work. Nonetheless, there are two 
parameters of crucial importance for any metric of the proposed form. The 
farthest point $r_{\rm m}$ reached by the non-standard transient region and the 
exact duration of the bouncing $t_{\rm b}$. 

On the one hand, let us stress again that $r_{\rm m}$ is always beyond the 
gravitational radius. In our description we have taken $r_{\rm m}$ to be 
smaller than $r_{\rm i}$ but in principle it could be even larger than $r_{\rm 
i}$. On the other hand, we have regularized the singularity in such a way that the 
total duration of the entire bounce is exactly twice the classical collapsing 
time. In more realistic terms we expect the total duration to be somewhat 
larger than this but of the same order. 

To conclude, the entire process has nothing to do with a regular black hole 
evaporation. On the one hand, the appearance and disappearance of trapped 
regions is so rapid that the process cannot be considered as an evaporation; it 
has no relation whatsoever with Hawking evaporation. On the other hand, since 
as a first approximation there has been no dissipation, the stellar system has 
come back to its initial configuration before the collapse.

\subsection{New equilibrium possibilities}
\label{Subsec:equilibrium}

If there exists a regularization of the classical standard general relativity 
behaviour of the form we have described, the collapse process itself would not 
constitute the final stage of collapse in stellar physics. One 
would immediately be impelled to wonder about what would happen after the bounce. The 
search for new states of equilibrium on the one hand and the understanding of 
the transient collapse process itself on the other become entirely distinct 
issues.

In an ideal situation, perfectly spherically symmetric and without dissipation, 
the collapsing body would enter into a never-ending cycle of contracting and 
expanding phases. In a realistic situation, though, one expects that the system 
will dissipate at least quantum-mechanically while searching for new 
equilibrium configurations. Here the panorama of possibilities is almost 
unexplored. Let us discuss briefly the two possibilities we think more 
plausible.

In~\cite{Barcelo2008} it was shown that if the velocity of trapping-horizon 
crossing were rather small, then quantum effects of vacuum polarization would 
become so powerful that they might even stop the collapse. It is very unlikely 
that the expected almost-free-falling collapse of stellar structures like 
neutron stars would lead directly to strong vacuum polarization effects. 
However, in our scenario, when taking into account dissipation, one would expect 
that each new recollapsing phase would start from a position closer to its 
Schwarzschild radius than the previous one. In this way, at some stage vacuum 
polarization effects could start to be relevant and even stop further 
collapses. This might lead to hypothetical almost-stationary structures 
hovering extremely close but strictly outside their gravitational radius. 

There might be other mechanisms underlying the stabilization of stellar 
structures close to their gravitational radius. What is relevant here is that 
the final metastable object could be small, dark, and heavy, but without black- 
or white-hole districts (see~\cite{Visser2008} and references therein). These 
black stars will not be hollows in space, they will be filled with matter (they might also be gravastars \cite{Mazur2001}). And 
most importantly, since they have no horizons they will in principle be completely open to 
astrophysical exploration. Classical general relativity black holes might still continue being very good models for the external gravitational behaviour of these black stars.

What would happen with Hawking radiation and Hawking evaporation in this 
scenario? During the transient phases one would expect quantum dissipation in 
the form of particle production. This particle production will be in general 
non-thermal, though at trapping-horizon crossings it would have the form of 
short bursts of thermal radiation~\cite{Stephens1994,Barcelo2006,Barbado2011}. 
When the system becomes stabilized close but outside its Schwarzschild radius, 
it might emit or not and with different spectral properties depending on the 
specifics of the structure, which at this stage are difficult to envisage. 
However, it is interesting to realize that there are at least two ways in which 
these objects could acquire emission and evaporation properties resembling 
Hawking's scenario. One such structure could emit a Hawking-like flux if it 
were continuously and asymptotically approaching its Schwarzschild 
radius~\cite{Barcelo2006} or if it were pulsating in a close to free-fall 
manner~\cite{Barbado2011}. Should this radiation exist, it would in both 
cases be Planckian but not strictly thermal, as correlations are maintained by 
both mechanisms. In fact, these scenarios do not invite us to wonder whether 
information is lost or not, as no singularities and no long-lived trapping 
horizons are formed in the first place.

The other possibility is that somehow a strong dissipative effect leads to a 
stationary structure of high density surrounded by a trapping horizon. This 
object should be subject to Hawking evaporation and in fact would be nothing 
but a REBH. Again, they will suffer from the lifetime problem, but in an 
ameliorated form as the physics of the quick transient bounce would have shown 
part of their nature.

\subsection{An underlying hypothesis}
\label{Subsec:hypothesis}

The past thick dashed line in Fig.~\ref{Fig:bh-wh-diagram} marks the boundary where non-standard gravitational 
effects start to happen. It is born at $r=0$ and travels outwards even though 
the light cones are pointing inwards. This signal, should it exist, cannot 
follow the causality associated with the gravitational light cones. Rather it 
must follow an underlying causality that is explored only when Planck energies 
are at stake. This background causality should be non-dynamical and trivial in the sense of containing no horizons whatsoever, the simplest 
example one can think of being a Minkowskian structure. This is the only 
assumption we need for our proposal to make sense.

Otherwise general-relativistic light cones could not suffer such a dramatic turn. The only possibility apparently left would be that the light cones did not quickly reverse its tilting but only slowly recover their unbent positions before collapse. The geometry would be the one described for instance in~\cite{Roman1983} which is qualitatively equal to that in Rovelli-Vidotto's evaporation model. They consider the slow growing (as seen from outside) or anti-evaporation of the internal horizon as a kind of quantum bounce.

Our assumption can be motivated both from particle physics and 
condensed matter physics. On the one hand, our proposal connects naturally with 
Rosen's reformulation of general relativity as a nonlinear theory on a flat 
Minkowski background \cite{Rosen1940}. This reformulation indeed goes further 
than the standard formulation of general relativity in the sense that it is a 
convenient effective framework to describe the switching-off of gravity at high 
energies. Rosen's reformulation can be understood as the long-wavelength limit 
of a nonlinear theory of gravitons (see \cite{Barcelo2014} and references 
therein). It is still an open possibility that an ultraviolet completion of 
such a theory would exhibit asymptotic freedom (as its QCD cousin). On the 
other hand, similar ideas also appear when thinking of gravity as an emergent 
notion in a condensed matter framework~\cite{Barcelo2010a} (see also~\cite{Jannes2009}). The nonlinear 
theory of gravity describes in that case the behaviour of collective degrees of 
freedom. There, it is reasonable to think that the first quantum gravitational 
effect is that, above some Planckian energy scale, the collective degrees of 
freedom corresponding to gravity are diluted, leaving a Minkowskian background 
for the matter excitations. 

Let us detail how the switching-off of gravity can be described. The action of 
a theory of this sort should be expressible in terms of a composite field 
$g_{ab}=\eta_{ab}+\lambda h_{ab}$ , where $\eta_{ab}$ is the flat background 
metric, $h_{ab}$ the graviton field and $\lambda$ the coupling constant which 
controls the nonlinearities of the gravitational sector as well as the coupling 
to matter (take Rosen's formulation as an example, see~\cite{Barcelo2014}). For 
any nonzero value of $\lambda$ the field equations of this theory are 
equivalent to the Einstein field equations for a metric $g_{ab}$. Thus in this 
regime one recovers general relativity, with a (generally curved) metric 
$g_{ab}$ controlling the effective causality of the spacetime. However, the 
structure of the theory permits us to consider the limit $\lambda\rightarrow0$, 
in which the nonlinearities disappear and matter is effectively decoupled from 
the graviton field, which evolves separately as a free field. In this limit the 
causality of the spacetime, given by $\eta_{ab}$, is no longer dynamical. The 
two conceptual frameworks above suggest that when high-energy phenomena are 
involved, this limit is indeed reached so that the underlying causality in the 
system, which is Minkowskian with no horizons whatsoever, is unveiled.

This view prioritizes the role of matter with respect to the dynamical causal 
structure contained in $g_{ab}$, and brings the 
general-relativistic and Newtonian descriptions of matter scattering in the 
presence of gravity, discussed in Subsec.~\ref{Subsec:white-hole}, a step closer to each other. To realize 
this one only needs to think about what happens in a local region around the 
distribution of matter undergoing gravitational collapse \cite{Barcelo2010a}. 
At 
some point it will enter the regime in which the local causal structure is 
Minkowskian and there is no trace of gravity. After a scattering process which 
takes place in the absence of gravity and which can be idealized as a first 
approximation as dissipationless, the lump of matter will effectively bounce back, now expanding in 
time. Notice that it is not needed to consider that some kind of repulsive 
force 
acts in this regime, and that its existence would only lead to quantitative 
changes in this picture. If we keep following the expanding distribution of 
matter we will exit the high-energy regime in which the causal structure is not 
dynamical and the usual general-relativistic picture with a gravitational field 
$g_{ab}$ will be restored. However, this dynamical causal structure is a 
secondary character in this story, so that it will naturally adapt itself to 
the 
distribution of matter in spacetime, with the corresponding lightcones pointing 
outwards. This leads to the global geometry we are discussing in detail in this 
paper. 

The gravitational switching-off and -on process we have described would have an 
interpretation in general relativity in terms of an effective energy content 
that violates certain energy conditions in some regions. We advance here the following 
interpretation of the ideal dissipationless bouncing process. When the 
collapsing structure reaches Planckian densities it acquires a ``quantum 
modification'' leading to an effective density $\rho-\rho_{\rm q}$. At some 
point this effective density becomes negative and thus repulsive. This negative 
density is compensated by a burst of positive energy that is expelled out of 
the structure through the underlying causality (this is what the bottom thick dashed 
line in Fig.~\ref{Fig:bh-wh-diagram} intends to represent). This positive energy 
burst reaches some point $r_{\rm m}$ outside the gravitational radius. The 
time-reversed process should not be understood as a sort of collapse of this 
energy burst, but rather as describing the attenuation of the effect of its 
propagation through the dynamical geometry.

\section{Observational signatures}
\label{Sec:observational}

If our idea is at work in Nature it will have many observable effects. As we 
said above, in the collapse of, say, a neutron star, matter would remain 
apparently frozen at the Schwarzschild radius for just a few tenths of a 
millisecond before being expelled again. In realistic situations the bounce 
will not be completely time-symmetric: part of the matter will go through 
towards infinity in the form of dissipative winds while another part would tend 
to recollapse. In this way one would have a brief and violent transient phase, 
composed of several bounces, followed by the formation of a new (metastable) 
equilibrium object with the remaining mass. 

At this stage we can just speculate about the possible signatures of both 
phases. As we have described, in our proposal, even neglecting dissipation, the 
metric is non-Schwarzschild during a short time interval in a region (the dark grey 
triangle in Fig.~\ref{Fig:bh-wh-diagram}) extending outside the gravitational 
radius. 
The farthest point of this region, $r_{\rm m}$, is an unknown parameter of the 
geometry. In the effective energy description it corresponds to the farthest 
point reached by the energy burst created when the collapsing matter reaches 
Planckian densities. Any observer monitoring the collapse process from an 
orbit with radius 
smaller than this radius will clearly gravitationally perceive the burst 
crossings even before he observes the reappearance of the collapsed body. 
The fact that a region outside the horizon must be affected in order 
to construct a time-symmetric bounce is highly interesting. It is a general 
belief that corrections to general relativity should only be important in 
regions of spacetime in which the curvature is high enough. The near-horizon 
region of stellar black holes does not fulfill this requirement, so that the 
behaviour of the gravitational field in these regions is not regarded as 
suitable to be drastically modified. While this theoretical prejudice must be 
confronted with experiments, a more detailed study of the observational 
implications of the solution to the lifetime problem we propose here could lead 
to indirect traces of this ``heretic'' behaviour.

When considering realistic situations with dissipation, the transient phase 
might leave some traces, for instance in the physics of Gamma Ray Bursts: One 
would expect some signatures associated with a reverberant collapse. However, 
in the collapsar model of GRBs (see {\it e.g.}~\cite{MacFadyen1998}) the emission 
zone is supposed to be very far from the collapsed core. This means that the 
connection between the processes at the core and those at the external wind 
shells could be very far from direct.  

Regarding the new equilibrium phase, there might be several observational 
opportunities if the object happens to be a black star ({\it i.e.} a stellar 
object hovering outside its gravitational radius). In this case one has to 
distinguish between light coming from processes occurring at the surface of the 
object and light produced far from the object but that becomes reflected at its 
surface. Any process occurring at the surface will be hardly seen from outside 
due to the tremendous redshift factor that these objects could produce near 
their surface. 
However, if one sent a radar signal straight towards a black star the light 
would first be blue-shifted and, in the case of a perfect elastic scattering at 
the surface, it would come back with precisely the same initial frequency, 
picking up no red-shift at all. It would just pick up a time-delay factor 
because of the time spent in the surroundings of the surface. The presence 
of this delay factor is what would primarily distinguish a black star from a 
black hole: a black hole leads to no echo. As already emphasized 
in~\cite{Barcelo2010a}, the crucial point is that the delay expected from one 
of these black stars is not large at all (nothing comparable with the age of 
the universe). If this were not the case, there would be no practical way to 
distinguish between both objects with radars. The general-relativistic delay 
can be calculated to be 
\begin{equation}
T=2\int_{r_s}^{r_0}\frac{dr}{1-2M/r}=2\left[r_0-r_s+2M\ln\left(\frac{r_0-2M}{r_s-2M}
\right)
\right],
\label{delay}
\end{equation}
with $r_0$ the observation point. The divergent term is logarithmic so that it 
can never become too large in realistic situations. For instance, for a Solar 
mass black star with a radius larger than its Schwarzschild radius ($3\times 
10^3$ m) by the tiny amount of $10^{-75}\ \mbox{m}$ (which is about $10^{-40}$ 
times the Planck length), a radar signal sent from a distance of 8 
light-minutes 
would acquire a gravitational delay of only a few milliseconds and would echo 
back after about 16 minutes plus few milliseconds, in sharp contrast with the 
infinite amount of time necessary in the case of a proper black hole.

\section{Summary and conclusions}
\label{Sec:summary}

The fate of high-density matter undergoing gravitational collapse is a 
fascinating problem which has been in the spotlight for many years. The black 
hole information-loss paradox is the quintessential example of the issues one 
faces when trying to combine general relativity and quantum mechanics to get 
some hint about this process. Here, however, we wanted to make a change of 
perspective and highlight what we find an unphysical feature: the 
characteristic dynamical time scales associated with REBHs are ridiculously large in comparison to other scales in Nature. This observation alone motivates the search for radical alternatives to the standard picture of gravitational collapse. 
Theories of emergent gravity naturally suggest a refreshing alternative, first introduced in [16] and briefly studied in [18]: the standard collapse to form an object essentially indistinguishable from a black hole is substituted by a perfectly time-symmetric collapse and bounce (in the idealized case of a spherical distribution of matter starting from a finite radius and in the absence of classical and quantum dissipation). 
In this paper we further develop the geometric properties of this alternative. There exists an entire family of geometries, with functional and parametric
degrees of freedom which at the moment cannot be fixed without knowing additional features of the underlying theory. However, the most relevant details are robust with respect to these variations, and can be summarized in two points:
\begin{itemize}
\item
The time lapse associated with the collapse is very short (of the order of milliseconds
for neutron-star-like initial configurations) for geometries which simply regularize
the behaviour of the collapse near the singularity. This is equally true both for
observers attached to the structure as well as for distant stationary observers.

\item
It is mandatory that a certain open region outside the Schwarzschild radius deviates
from the usual static and spherically symmetric solution. This deviation as well
as the matter bounces will lead to characteristic imprints in the transient phase.
While a detailed study must be done in order to find experimental signatures
of these phenomena, the short timescales associated with all these processes are
encouraging.
\end{itemize}

The plausibility of the time-symmetric bounce process rests on the assumption that the general-relativistic description of gravity is not fundamental, so that modifications to the dynamical behaviour of the spatiotemporal causal structure at high energies are conceivable. The standard view is that ``quantum corrections'' to a general-relativistic geometry could only occur in high curvature regions. Our discussion points out that this assertion contains at least one assumption that is typically unstated: The non-existence of a deeper causality of Minkowskian character (that is, with no horizons). If this causality exists, it is not difficult to imagine that the echo of the high-energy collision produced in the location where the classical singularity was supposed to appear would be transmitted outside through the deeper (high-energy) causality, and modify the effective light-cone structure (i.e. the light-cone structure of general relativity) even in places with very small effective curvature. One also has to take into account that the non-general-relativistic modification of the effective light-cone structure we speak about occurs just in a brief transient region, being the manifestation of the rapid process of switching off and on of gravity. The existence of this deeper causality is indeed a necessary and sufficient assumption, thus making a tight connection between certain properties of the high-energy behaviour of the gravitational interaction and low-energy solutions which do not present the lifetime problem.

This rapid bounce alternative radically changes the discussion of the possible endpoints of gravitational collapse. In particular, it makes plausible that the final object is a compact object with no horizons whatsoever. As with other compact objects in Nature, the structure of the transient would be given by ``dirty'' (non-geometrical) physics whose details still need to be filled out. If this view is indeed realized in Nature, black holes could just be  an idealized approximation to the ultimate stationary objects.

With respect to future work, we can foresee three main lines of research. 
First, the search for possible new states of stellar 
equilibrium becomes most interesting in the light of the present paradigm. Second, this proposal would 
find strong theoretical support if one could develop a better understanding of theories or 
frameworks beyond classical general relativity leading to the described 
behaviour. Finally, from an observational point of view, one should look at
high-energy phenomena associated with gravitational collapse of massive bodies, 
such as GRBs, for potential signatures of the main qualitative features of our 
proposal such as the reverberant collapses and short timescales. While plenty 
of work remains to be done, the possibility of testing fundamental properties 
of gravitation in astrophysical experiments is always tantalizing, especially 
when there exists the possibility that high-energy gravitational physics is not 
resigned to be oppressed by the structure of long-lived trapping horizons.

\section{Epilogue}
\label{Sec:epilogue}
The geometry proposed by Haggard-Rovelli~\cite{Haggard2014} is causally the 
same we described in~\cite{Barcelo2010a} and in more detail in the 
essay~\cite{Barcelo2014e} and this paper. As far as we can see the only but 
crucial difference between the two geometries is precisely the duration of the 
bounce as seen by external observers far from the collapsing object, which is 
related to the different underlying hypotheses in both cases. The 
Haggard-Rovelli proposal still exhibits the lifetime problem, if only 
slightly alleviated because the bouncing time scales with $m^2$ and not with 
$m^3$ as the standard black-hole evaporation lifetime. The reason is that in 
the Haggard-Rovelli scenario it takes a long time for the quantum effects to 
accumulate so as to produce a significant deviation from the classical general 
relativity solution. The process could be described as a quantum tunneling 
between two classical solutions with a very low probability or very long 
lifetime. This is why the process might not need to invoke any violation of 
the dynamical general-relativistic causality but keep it as a fundamental notion.

In our proposal, on the contrary, the bouncing time is twice the proper 
collapsing time, and thus a very short time, of the order of tenths of a 
millisecond for neutron-star-like objects. This is possible as long as one 
accepts a modification of general relativity such that the causal structure of 
spacetime is non-standard in a transient region. For instance, this behaviour 
could be achieved with an 
underlying structure to general relativity in which gravity can be switched off 
during brief periods of time while still leaving a Minkowskian background causality. 
This assumption has permitted us to construct a solution which is free from the 
lifetime problem. As we have discussed, the fact that this solution cannot 
be reached in a framework in which the causality of general relativity plays a 
fundamental role (such as in Haggard-Rovelli's construction) opens an 
interesting possibility of discriminating between radically different 
behaviours of gravity at high energies in near-future astrophysical 
experiments. 

It is also interesting to realize that the same logarithmic behaviour for the 
delay of a signal which is reflected at a surface close to the horizon 
(Eq.~(\ref{delay}) here and Eq.~(8) in~\cite{Haggard2014}) is used to argue in 
almost opposite directions in both articles. Haggard-Rovelli use it to argue 
that one could obtain very long external delays; we use it to highlight that, 
being logarithmic, even in situations in which the reflection is produced 
extremely close to the horizon, the delay will not be large.

\ack

Financial support was provided by the Spanish MICINN through the projects 
FIS2011-30145-C03-01 and FIS2011-30145-C03-02 (with FEDER contribution), and by 
the Junta de Andaluc\'{\i}a through the project FQM219. R. C-R. acknowledges 
support from CSIC through the JAE-predoc program, cofunded by FSE.

\section*{References}

\bibliographystyle{unsrt}
\bibliography{wh-proposal_rev}

\end{document}